\documentstyle[12pt,epsf]{article}
\newcommand{\beq}{\begin{equation}}
\newcommand{\eeq}{\end{equation}}
\newcommand{\beqa}{\begin{eqnarray}}
\newcommand{\eeqa}{\end{eqnarray}}
\newcommand{\beqan}{\begin{eqnarray*}}
\newcommand{\eeqan}{\end{eqnarray*}}

\newcommand{\ben}{\begin{enumerate}}
\newcommand{\een}{\end{enumerate}}
\newcommand{\bfl}{\begin{flushleft}}
\newcommand{\efl}{\end{flushleft}}
\newcommand{\ba}{\begin{array}}
\newcommand{\ea}{\end{array}}
\newcommand{\btab}{\begin{tabular}}
\newcommand{\etab}{\end{tabular}}
\newcommand{\bit}{\begin{itemize}}
\newcommand{\eit}{\end{itemize}}

\newcommand{\hs}{\hspace}

\def \g5 {\gamma_{5}}

\newcommand{\prepr}[1] {\begin{flushright} {\bf #1} \end{flushright} \vskip
1.5cm}
\newcommand{\titul}[1] {\begin{center}{\Large {\bf #1 } } \end{center} \vskip 1.cm}

\newcounter{muni}

\begin{document}
\vspace{.1cm}
\hbadness=10000
\pagenumbering{arabic}
\begin{titlepage}
\prepr{Preprint hep-ph/xxxxx\\PAR/LPTHE/98-45\\ September 1998 }
\titul{Is the lepton flavor changing observable in Z$\to \mu^{\mp} + \tau^{\pm}$ decay ?}
\vspace{5mm}
\begin{center}

{\bf Xuan-Yem Pham \footnote{\rm Postal address: LPTHE, 
Tour 16, $1^{er}$ Etage, 
4 Place Jussieu, F-75252 Paris CEDEX 05, France. \\
. \hspace{5mm} Electronic address : pham@lpthe.jussieu.fr }
}
\end{center} 
\vspace{0.5mm}
\begin{center}
{\large \bf \it
Universit\'e Pierre {\it \&} Marie Curie, Paris VI \\
Universit\'e Denis Diderot, Paris VII \\
Physique Th\'eorique et Hautes Energies
} 
\end{center}

\thispagestyle{empty}
\vspace{0.5mm}
\hspace{0.01cm} \large{} {\bf  \hspace{0.1mm} Abstract-}    
\normalsize
Neutrino oscillations as suggested by the recent Super-Kamiokande result implies that lepton numbers could be violated, and  Z$\to \mu^{\mp} +\tau^{\pm}$ is a typical manifestation. We point out that for this mode, the GIM cancelation is much milder with only a logarithmic behavior $\log (m_3 /m_2)$ where $m_j$ is the neutrino mass. The ratio $\Gamma$(Z $\to \mu^{\mp} + \tau^{\pm})/\Gamma($Z $ \to \mu^{-} + \mu^{+})$ could be about $10^{-5}-10^{-6}$, in sharp contrast with the vanishingly small rate $\tau^{\pm}\to \mu^{\pm} 
+ \gamma$ strongly
suppressed by a quadratic power  $(m_3^2-m_2^2)/ M_{\rm W}^2$. Being complementary to neutrino oscillation experiments, measurements of Z$\to \mu^{\mp} +\tau^{\pm}$ -- which are at hand with the present colliders -- would give one more constraint to the lepton mixing angle $\sin 2\theta_{jk}$ and the neutrino mass ratio $m_j/m_k$.

\vspace{10mm}
{\bf PACS numbers : 12.15.Ff, 12.15.Lk, 13.38.Dg, 14.60.Lm, 14.60.Pq }
\end{titlepage}

\newpage

Evidence for the transmutation between the two neutrinos species $\nu_\mu \leftrightarrow \nu_\tau$ is recently reported  by the Super-Kamiokande collaboration$^{(1)}$. As a consequence, neutrinos could have nondegenerate tiny masses and lepton numbers would no longer be conserved. Hence, besides the well known neutrino oscillations phenomena, the decays $\tau^{\pm} \to \mu^{\pm} + \gamma$, Z$ \to \mu^{\mp} +\tau^{\pm}$ could occur. 

The purpose of this note is to point out that, contrarily to the hopelessly small$^{(2)}$ branching ratio B$(\tau^{\pm} \to \mu^{\pm} +\gamma) \approx 10^{-40}$, the typical lepton flavor changing process Z$ \to \mu^{\mp} +\tau^{\pm}$ could be 
${\it experimentally}$  ${\it accessible}$. The ratio of branching fractions 
B(Z$\to \mu^{\mp} +
\tau^{\pm})$/B(Z $\to \mu^{-} + \mu^{+})$ is estimated to be about 
$ 10^{-5} - 10^{-6}$ which is much larger than naively expected. With ten millions of Z produced per year at the present colliders, its measurement is therefore at hand. The interest of  this decay is obvious since it is complementary to neutrino oscillation experiments, and will give one more information on the lepton mixing angle $\theta_{ij}$ as well as  the ratio $m_j/m_k$.

Similarly to the CKM flavor mixing in the quark sector, the neutrino gauge-interaction eigenstates $\nu_{\rm e}$, $\nu_\mu$ and $\nu_\tau$ are linear combinations of the three neutrino mass eigenstates $\nu_1$, $\nu_2$ and $\nu_3$ of nonzero and nondegenerate masses $m_1$, $m_2$ and $m_3$ respectively. Thus
\begin{equation}
\pmatrix{ \nu_{\rm e}\cr \nu_\mu\cr \nu_\tau\cr}
=\pmatrix { U_{{\rm e}1} & U_{{\rm e}2} & U_{{\rm e}3}\cr
                 U_{\mu 1} & U_{\mu 2 } & U_{\mu 3}\cr
                 U_{\tau 1}  & U_{\tau 2} & U_{\tau 3}\cr}
               \pmatrix {\nu_1\cr \nu_2 \cr \nu_3\cr} 
\equiv \; {\cal U}_{\rm lep}\;\pmatrix {\nu_1\cr \nu_2 \cr \nu_3\cr } , 
\label{eq:1} 
\end{equation}
where the $3\times 3$ matrix $\;{\cal U}_{\rm lep}\;$ is unitarity.

The weak interaction effective Lagrangian for charged current of leptons can be written as
$${\cal L}_{\rm eff} ={{\rm G}_{\rm F} \over \sqrt{2}} L^{\dagger}_\lambda L_\lambda \;,$$ 
where the charged current $L_\lambda$ is
$$L_\lambda =\sum_{j =1}^{3} \overline{\ell}\gamma_{\lambda}(1-\gamma_5) \nu_j U_{\ell j}\;.$$
Here $\ell$ stands for e$^-$, $\mu^-$, $\tau^-$ and $\nu_j$ (with $j=1,2,3$) are the three neutrino mass eigenstates. For any fixed $\ell$, one has $\sum_{j}\vert U_{\ell j}\vert^2 =1$. For instance the $\nu_\mu$ operationally defined to be the invisible particle missing in the $\pi^+ \to \mu^{+} + \nu_{\mu}$ is initially a superposition of $\nu_1$, $\nu_2$ and $\nu_3$, in the same way as the K$^0$ meson produced by strong interaction, say by $\pi^- +{\rm p} \to {\rm K}^0 +\Lambda$, is initially a superposition of the mass eigenstates K$^0_{\rm L}$ and K$^0_{\rm S}$ with nondegenerate masses $m_{\rm L} \not= m_{\rm S}$.

 In the most general renormalizable R$_{\xi}$ gauge, at one loop level to order $g^3$ -- where $g=e/\sin \theta_{\rm W}$ is the weak interaction coupling constant -- there are in all ten Feynman diagrams that contribute to the decay Z$ \to \mu^{\mp} +\tau^{\pm}$. Three of them are depicted in Figs. 1--3.
The seven others, not shown here, are similar to Figs. 1--3 in which 
the W$^{\mp}$ are replaced in all possible ways by the "would be" Goldstone
 bosons $\Phi^\mp$, those absorbed by the gauge bosons W$^{\mp}$  to render them massive by the Higgs mechanism.

A careful examination of these diagrams shows that Fig.1 provides by far the dominant contribution to the decay amplitude for which the logarithmic GIM 
suppression actually emerges, while the contributions of all other diagrams are power suppressed and vanishingly small. The principal reason is that we are dealing in Fig.1 with two propagators of nearly massless fermions. Due to the unitarity of the ${\cal U}_{\rm lep}$ reflecting the GIM cancelation mechanism, the  divergence as well as the $m_j$-independent finite part of the loop integral do not contribute to the decay amplitude because they are multiplied by $\sum_j (U_{\mu j}^* U_{\tau j}) =0$. Only the $m_j$-dependent finite part of the loop integral is relevant. The finite part could be easily guessed by approximating the W propagator with $\;$ i$/ M_{\rm W}^2$, the W mass plays the role of the loop integral
 momentum cutoff. Hence Fig.1 looks like the familiar vacuum polarization in Fig.1bis from which appears the standard  
$\log (q^2 /m_j^2)$ when $m_j^2\ll q^2$, $q^\mu$ being the four-momentum of the external gauge boson Z.
An exact calculation confirms this expectation.

If we write the Z$(q) \to \nu_j(p_1) +\overline{\nu}_j(p_2)$ amplitude as
$${-{\rm i} g\over 4 \cos \theta_{\rm W}}\; \varepsilon^{\mu} (q) \overline{u}(p_1) \gamma_\mu (1-\gamma_5) v(p_2)\;,$$%
then the Z$(q) \to \mu^-(p_1)+\tau^+(p_2)$ decay amplitude is found to be
$${\cal A} \times {-{\rm i} g\over 4 \cos \theta_{\rm W}} \;\varepsilon^{\mu} (q) \overline{u}(p_1) \gamma_\mu (1-\gamma_5) v(p_2)\;,$$%
where 
\begin{equation}
 {\cal A}= {1\over 3\pi^2}\left({-{\rm i} g \over 2\sqrt{2} \cos\theta_{\rm W}}\right)^2 \sum_jU_{\mu j}^* U_{\tau j} \log x_j \;,\;\;\; x_j = {m_j^2\over M_{\rm W}^2} \;.
\label{eq:2} 
\end{equation}
In (2), we only keep the dominant contribution $\log x_j$. Terms like $x_j$, $x_j\log x_j$, $x_j^2 \log x_j$ and dilogarithms  (Spence functions) of $x_j$ 
are completely negligible. The problem considered here is conceptually 
similar to the quark flavor changing s$\leftrightarrow $d mediated by a virtual 
Z exchange$^{(3)}$ which governs the rare decay K$\to \pi +\nu +\overline{\nu}$. However, since masses and momenta of the external particles outside the loops were neglected$^{(3)}$ -- with perfectly legitimate reasons -- the smoothly logarithmic factor (which multiplied by $q^2$, $q$ being the virtual Z momentum taken to be zero) cannot be revealed$^{(3)}$.  For the external real Z discussed here, this approximation $q^2=0$ obviously cannot be applied, and the factor $\log x_j$ emerges. In fact, the vacuum polarization $\Pi(q^2)$ mentioned above behaves$^{(4)}$ like $q^2/ m_j^2$ for $q^2\ll m_j^2$, and like $\log (q^2/m_j^2)$ for $m_j^2\ll q^2$, $m_j$ being the internal fermion mass. Quark flavor changing in Z decay has been extensively studied in the literature$^{(5)}$. 

We have from (2)
\begin{equation}
 {\Gamma({\rm Z} \to \mu^{\mp}+\tau^{\pm}) \over \Gamma({\rm Z} \to \mu^- +\mu^+)}= {\alpha^2\over 36\pi^2 \sin^4\theta_{\rm W}\cos^4\theta_{\rm W} }\left| {\cal B}\right|^2 \;,
\label{eq:3} 
\end{equation} 
where 
\begin{equation}
{\cal B}=
\sum_{
j=2}^3 U_{\mu j}^* U_{\tau j} \log \left({m^2_j\over m^2_1}\right) \;.
\label{eq:4} 
\end{equation}
From (2) to (4), we have used $\sum_j U_{\mu j}^* U_{\tau j}=0$ to get rid of  the first $\nu_1$ mixing parameter
$U_{\mu 1}^* U_{\tau 1}$.
The factor ${\cal B} $ in (4)
tells us that when the $m_j$ are degenerate, the lepton flavor mixing does not occur and the decay Z $ \to \mu^{\mp} + \tau^{\pm}$ identically vanishes.

To estimate  ${\cal B}$, let us assume$^{(6)}$ the following form of the ${\cal U}_{\rm lep}$, neglecting CP violation in the lepton sector: 
\begin{equation}
{\cal U}_{\rm lep} = \pmatrix{ \cos\theta_{12} & -\sin\theta_{12}& 0 \cr 
{1\over \sqrt{2}}\sin\theta_{12} & {1\over \sqrt{2}}\cos\theta_{12}  & 
{-1\over \sqrt{2}}\cr
        {1\over \sqrt{2}}\sin\theta_{12}&{1\over \sqrt{2}}\cos\theta_{12}  & 
{1\over \sqrt{2}}\cr}\,.        
\label{eq:4} 
\end{equation}
The mixing angle $\theta_{23}\approx 45^0$ is suggested by the Super-Kamiokande data and
the $\theta_{13}\approx 0^0$ comes from the Chooz data$^{(6)}$ which give $\theta_{13} \leq 13^0$, whereas $\theta_{12}$ being arbitrary. Although  $\theta_{12}$ is 
likely small $\approx 0^0$, however the maximal mixing $\theta_{12}\approx 45^0$ may be also possible allowing $\nu_{\rm e}\leftrightarrow \nu_\mu$ (as suggested by the LSND experiment). Taking $\theta_{12}$ in the range 
$0^0$--$45^0$, and using$^{(1,6)}$ $\Delta m^2_{23} =\vert m_3^2-m_2^2\vert =2\times 10^{-3}$ eV$^2$, $m_3 \geq 5\times 10^{-2}$ eV, then $\vert{\cal B}\vert^2$ is of order of unity 
[$\vert{\cal B}\vert^2 \sim {\cal O}(1)$], and we get
\begin{equation}
 {\Gamma({\rm Z}\to \mu^{\mp} +\tau^{\pm})\over \Gamma ({\rm Z} \to \mu^{-} +\mu^{+})} \approx 10^{-6} - 10^{-5} \;.     
\label{eq:5} 
\end{equation}

 We note that the low energy e$^{+} +{\rm e}^{-} \to \mu^{\mp} + \tau^{\pm}$ reaction might be also observable, however compared to the one-photon exchange $\sigma$(e$^{+} +{\rm e}^{-} \to \mu^{+} + \mu^{-})$ cross-section, the  $\sigma ({\rm e}^{+} +{\rm e}^{-} \to \mu^{\mp} + \tau^{\pm})$ is damped, besides the coefficient $|{\cal A}|^2$, by an additional $s^2/(M_{\rm Z}^2-s)^2$ multiplicative factor due to the Z propagator.

Independently of the precise numerical value of  ${\cal B}$, the experimental search of the rare decay mode Z $\to \mu^{\mp} +\tau^{\pm}$ is interesting on its own right for two reasons. First, although being higher order loop effect, the branching ratio is remarkably not negligible due to the smoothly logarithmic suppression. Second, while neutrino oscillations only provide the mass difference $\Delta m_{jk}^2$, lepton flavor changing Z decays give the ratio $m_j/m_k$. When combining these two processes, the absolute value of the neutrino mass $m_j$ can be obtained. Both reactions are mutually complementary in the determination of the neutrino masses and the lepton  mixing angles.  
\newpage



\large
{\bf Figure Captions}  :  \normalsize 

\begin{enumerate}
\item
Figures 1--3 : One-loop lepton flavor changing in Z decay.

\item
Figure 1bis : The same as Fig.1 in which the two $\mu$ and $\tau $ vertices are squeezed into a single four-fermion vertex. It looks like a bubble of the familiar Abelian gauge boson self energy.

\end{enumerate}

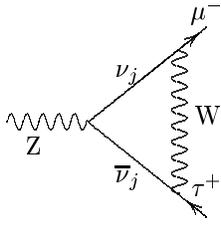
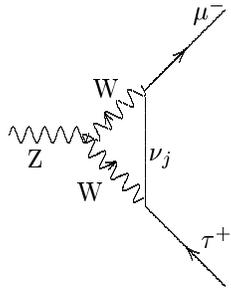
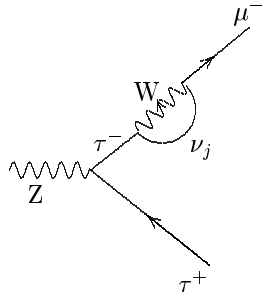
\end{document}